\documentclass[aps,prb,showpacs,twocolumn,groupedaddress]{revtex4-2}
\usepackage{amsmath}
\usepackage{cancel}
\usepackage{float}
\usepackage{ifsym}
\usepackage{color}
\usepackage[dvipsnames]{xcolor}
\usepackage{latexsym}
\usepackage{latexsym}
\usepackage{dcolumn}
\usepackage{epsfig}
\usepackage{amsfonts}
\usepackage{amsmath}
\usepackage{dsfont}
\usepackage{accents}
\usepackage[all,cmtip]{xy}
\usepackage{paralist}
\usepackage{enumerate}
\usepackage{amssymb}
\usepackage{bm}
\usepackage{graphicx}
\usepackage{wasysym}
\usepackage{hyperref}

\begin{document}

\title{\textcolor{blue}{Ferroelastic domain wall motion and collective domain switching in RbSCN}}
\author{V. Soprunyuk$^1$, A. Tr\"oster$^1$, J.~Pils$^2$, W.~Schranz$^2$}
\email{wilfried.schranz@univie.ac.at}
\affiliation{$^1$TU Wien, Institute of Materials Chemistry, 1060 Vienna, Austria}
\affiliation{$^2$University of Vienna, Faculty of Physics, Boltzmanngasse 5, A-1090 Wien, Austria}
\author{I. Rychetsky$^3$, A. Klic$^3$} 
\affiliation{$^3$Institute of Physics of the Czech Academy of Sciences, Na Slovance 2, 18200 Prague 8, Czech Republic}
\author{M.A. Carpenter$^4$} 
\affiliation{$^4$University of Cambridge, Departement of Earth Sciences, Downing Street CB2 3 EQ, Cambridge, U.K.}

\date{\today}

\begin{abstract}
Low frequency (0.05 - 40 Hz) dynamic elastic measurements and resonant ultrasound spectroscopy measurements (100-600 kHz) of RbSCN have been performed in the temperature region of the order-disorder improper ferroelastic phase transition at 
T$_c \approx$ 435~K. Quite similar to KSCN, the low frequency data show - in addition to the intrinsic phase transition anomalies - superelastic softening in a- and b-directions, resulting from movements of ferroelastic domain walls under dynamic stress. However, in contrast to KSCN, a sudden discontinuous increase of Young's modulus appears in RbSCN at 
{ T$^{\ast} < T_c $}, which is accompanied by a frequency dependent damping peak. 
This behaviour is reminiscent of a first order phase transition.\\
Heating RbSCN slightly above T$^{\ast}$, followed by subseqent cooling, removes all {signs of domain wall dynamics}. The results demonstrate, that the anomalies in RbSCN around $T^{\ast}$ result from collective domain switching events that are induced when the {temperature dependent critical pinning stress, $\sigma_c(T)$ falls below the  applied external stress $\sigma$, implying that $T^{\ast}(\sigma=\sigma_c)$.
This interpretation is supported by calculations of the temperature dependences of twin boundary widths $w$ and energies $F_w$, as well as the Peierls potential $V_0$ using a compressible pseudospin model, which leads to a critical pinning stress, $\sigma_c(T)$ that is in excellent agreement with experimental values of $T^{\ast}(\sigma_c)$. }
\end{abstract}


\maketitle

\section{Introduction}

Similar to other ferroics, ferroelastic materials form regions of symmetry related long range ordered states (domains), which are separated by domain walls  and which are switchable by the application of an external stress \cite{Salje1990}. Understanding the dynamical behaviour of domain walls (DWs), which is at the core of domain switching \cite{Gao2014}, is vital for the development of novel nano-technological devices \cite{Catalan2012, Meier2020}. Ferroelastic switching as a function of external stress often occurs via jerky movement of twin boundaries (TBs) \cite{Salje2010, Salje2011, Puchberger2017, Puchberger2018}, which can trigger larger collective switching events (avalanches) \cite{Nataf2020, Casals2021} forming a typical stress-strain hysteresis \cite{Salje2012}.

Quite an extensive literature on DW motion, domain switching and its influence on macroscopic susceptibilities exists \cite{Tagantsev2010}. Theoretical work includes Landau-free energy expansions in terms of a continuous order parameter appended with terms containing the order parameter gradient \cite{Sidorkin1997}, phase field calculations \cite{Bhattacharya2024, Bhattacharya2025} and other computer simulations \cite{Lu2019, He2022}, to name only a few of them. 
First experimental evidence that domain walls move in a jerky, collective way rather than smoothly, was presented already 1919 by H. Barkhausen \cite{Barkhausen1919} for magnetic systems. 
The idea of \textit{elastic} domain wall motion emerged gradually as experimental techniques improved and domain walls began to be treated as elastic interfaces moving in a disordered medium \cite{Zapperi1998, Paruch2013}.
The first clear experimental demonstration of elastic domain wall dynamics was published \cite{Lemerle1998} in the 1990s, showing the creep behaviour of a magnetic domain wall in an ultrathin Pt/Co/Pt film driven by a magnetic field.
Meanwhile DW dynamics have been studied in a number of ferroic (ferromagnetic, ferroelectric, multiferroic) systems, showing quite universal behaviour \cite{Kleemann2007} in agreement with theory.  
However, the ac response of \textit{elastic} DWs in \textit{ferroelastics} (proper or improper), and its relationship with magnetic and electric DW dynamics certainly merits  attention.

Here we show that crystals of the family MSCN (M=Rb, K) which exhibit an antiferrodistortive phase transition, where the order parameter is not the magnetization \textbf{M}, but given by a combination of pseudo-spin variables $S_i$ (i=1,2,3,4) coupled to strain $\varepsilon_{\alpha}$, exhibit quite interesting DW dynamics in response to a dynamically applied stress.

In Section \ref{sec:Phase Transitions in RbSCN} we review the main features of the phase transition properties of RbSCN. Section \ref{sec:dynamic elastic response} reports on  low (0.05 - 40 Hz) - and high (100 - 600 kHz) frequency elastic measurements of RbSCN. In Section \ref{sec:Discussion} we discuss the present data by combining well known theories of DW motion with a compressible pseudospin model. Section \ref{sec:Conclusion} concludes the paper.   

\section{Phase Transition in Rubidium Thiocyanate} \label{sec:Phase Transitions in RbSCN}

RbSCN crystals undergo an order-disorder improper ferroelastic phase transition (PT) at $T_c \approx$ 435 K from a high temperature tetragonal body-centred structure \cite{Shlyaykher2022} with space group $I4/mcm ~ (D_{4h}^{18})$, with 2 formula units in the primitive unit cell, to an ordered low temperature orthorhombic space group $Pbcm ~ (D_{2h}^{11})$ with 4 formula units in the unit cell (see Fig.1 of Ref. \onlinecite{Soprunyuk2025}).
Above $T_c$ the $SCN^-$-ions are head-tail disordered on average and below $T_c$ they order in an alternating arrangement, resulting in the loss of the centering translation $(\frac{1}{2},\frac{1}{2},\frac{1}{2})$.  
The lattice constants of the conventional tetragonal unit cell $\textbf{a}_t=(a_t,0,0)$,  $\textbf{b}_t=(0,a_t,0)$, $\textbf{c}_t=(0,0,c_t)$ of RbSCN \cite{Shlyaykher2022} and KSCN \cite{Yamada1963,Yamamoto1987} and of the orthorhombic unit cell,  are presented in Table~I.

\begin{table}
\caption{Lattice parameters of RbSCN and KSCN.}
\begin{tabular}{|c|c|c|}
\hline
 & RbSCN & KSCN \\
\hline
a$_t$ & 6.9198 ~\AA & 6.740~\AA  \\
c$_t$ & 8.2904 ~\AA & 7.832~\AA \\
\hline
a$_o$ & 6.8373~\AA & 6.676~\AA \\
b$_o$ & 6.8859~\AA & 6.691~\AA \\
c$_o$ & 8.0102~\AA & 7.606~\AA \\
\hline
\end{tabular}
\end{table}

Due to the symmetry reduction at the phase transition,  four domain states (DSs) \cite{Janovec1989,Schranz2019} are possible below T$_c$. DSs $1_1$ and $1_2$ as well as  $2_1$ and $2_2$  are related by lost translations $(\frac{1}{2},\frac{1}{2},\frac{1}{2})$ and are separated by antiphase boundaries (APBs), and DSs  $1_1$ and $2_1$ and $1_1$ and $2_2$ are related by 90$^{\circ}$ rotations, and are separated by ferroelastic DWs, also known as twin boundaries (TBs). 
The PT properties of RbSCN were previously studied in  detail by a number of experimental methods, including  NMR \cite{Blinc1995}, neutron scattering \cite{Blaschko1994}, Raman- and Infrared measurements \cite{Sathaiah1991}, as well as by molecular dynamics \cite{Hardy2000} and Monte Carlo \cite{Lodziana1999} simulations.

Thermodynamic properties \cite{Blinc1995,Blaschko1994} of the order-disorder PT in RbSCN have been well described by a compressible pseudospin model \cite{Schranz1989}, i.e. by including coupling between the pseudospin variables $S_i$ (i=1,..,4) and the strain components $\varepsilon_{\alpha}$ of the type $\alpha_{ijk}S_iS_j \epsilon_k$.  
These order parameter - strain couplings result in relatively large (dip like) anomalies of the longitudinal elastic constants, as previously measured in KSCN \cite{Schranz1994b,Schranz1994,Schranz2005}. 

Insights into the dynamical behaviour of the SCN fluctuations were gained by NMR measurements \cite{Blinc1995}. The fact that the $^{87}$Rb NMR line is homogeneously broadened, demonstrates that the disorder is dynamic. 
However, the time scale $\tau_{\eta}$ of the SCN head-tail fluctuations is so slow, that low symmetry $^{87}$Rb $\frac{1}{2} \rightarrow -\frac{1}{2}$ NMR angular rotation patterns are observed above $T_c$, implying that the high temperature phase of RbSCN consists of a large number of orthorhombic nanodomains with lifetimes $10^{-7} s < \tau_{\eta} < 10^{-3} s$.  Similar slow dynamics of OP fluctuations were also found in KSCN \cite{Blinc1991}.

Diffuse neutron scattering data of RbSCN \cite{Blaschko1994} showed that in the tetragonal phase the size $\xi$ of the nanodomains increases with decreasing temperature as 

\begin{equation}
\label{eq:xi}
\xi_a [\rm \AA] = \frac{55}{\sqrt{T-T_0}} \quad
\xi_c [\rm \AA] = \frac{38}{\sqrt{T-T_0}}
\end{equation}

Quite unusually, in the whole orthorhombic phase the correlation length is independent of temperature with $\xi_a \approx 30$~\AA  ~and $\xi_c \approx 25$~\AA. For comparison, in KSCN $\xi_a = 66/\sqrt{T-T_0}$, $\xi_c = 42/\sqrt{T-T_0}$ (at $T > T_c$) and  $\xi_a \approx \xi_c \approx 20$~\AA (at $T < T_c$)  was found \cite{Blaschko1991}.
Molecular dynamics calculations \cite{Parlinski1994,Lodziana1996,Blaschko1998} of a two dimensional KSCN model including order parameter - strain coupling of the type $\eta^2 \varepsilon$  reproduce this unusual behaviour. Later, it was argued \cite{Troster2000} that inhomogeneous strains suppress the growth of the clusters below T$_c$. Current MC simulations \cite{Troster2026} using a compressible pseudospin hamiltonian, including inhomogeneous strains, support this scenario.   

Cooling the crystals leads to a dense network of needle shaped ferroelastic domains. 
In the following we present recent dynamic elastic measurements, unveiling the dynamics of TBs of RbSCN, and compare them with results of KSCN \cite{Soprunyuk2025}.

\section{Dynamic elastic response of Rubidium Thiocyanate} \label{sec:dynamic elastic response}

{RbSCN single crystals were grown by slow evaporation from aqueous solution of NH$_4$SCN and Rb$_2$CO$_3$ by A. Fuith \cite{Fuith1997}. To prepare samples for TMA (thermomechanical) and DMA (dynamic mechanical)
measurements, the crystals were oriented (according to the known morphological habit planes), cut with a diamond saw and polished with diamond paste. The typical sizes of the samples were, prisms of L $\approx$ 3 mm and A$\approx$ 1.5 $\times$ 1.5 $mm^2$ for parallel plate measurements and bars of length  L $\approx$ 2-3 mm, thickness t$\approx$ 0.5 - 1 mm and width b$\approx$ 1 mm.}

\begin{figure}[h]
\centering
\includegraphics[scale=0.25]{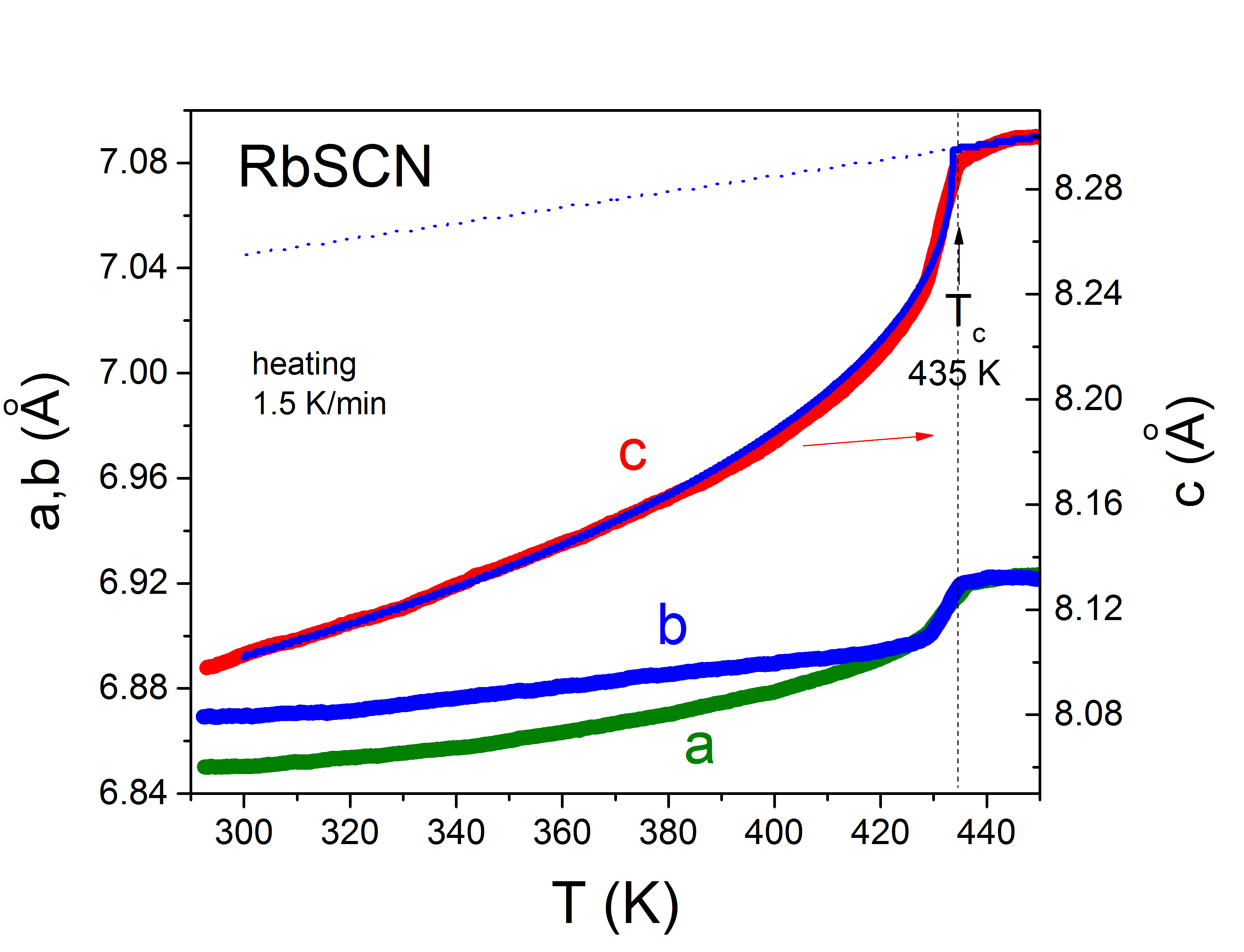}  
\caption{Thermal expansion of RbSCN in the three directions measured with TMA4000 (Perkin Elmer). The measured values of the sample dimensions $l_a, l_b, l_c$ are normalized to the lattice parameters a,b and c at T$_c$=435 K.}
\label{fig:thermal expansion}
\end{figure}

Fig.\ref{fig:thermal expansion} shows thermal expansion measurements of RbSCN using a TMA4000 (Perkin Elmer). The data are normalized to the lattice parameters \cite{Shlyaykher2022}  at 435~K to obtain the temperature dependencies of a, b and c.
The resulting room temperature lattice parameters agree reasonably well with those determined from X-ray scattering \citep{Shlyaykher2022}, {thereby confirming the assigned orientations of the samples}. 

\begin{figure}[h]
\centering
\includegraphics[scale=0.25]{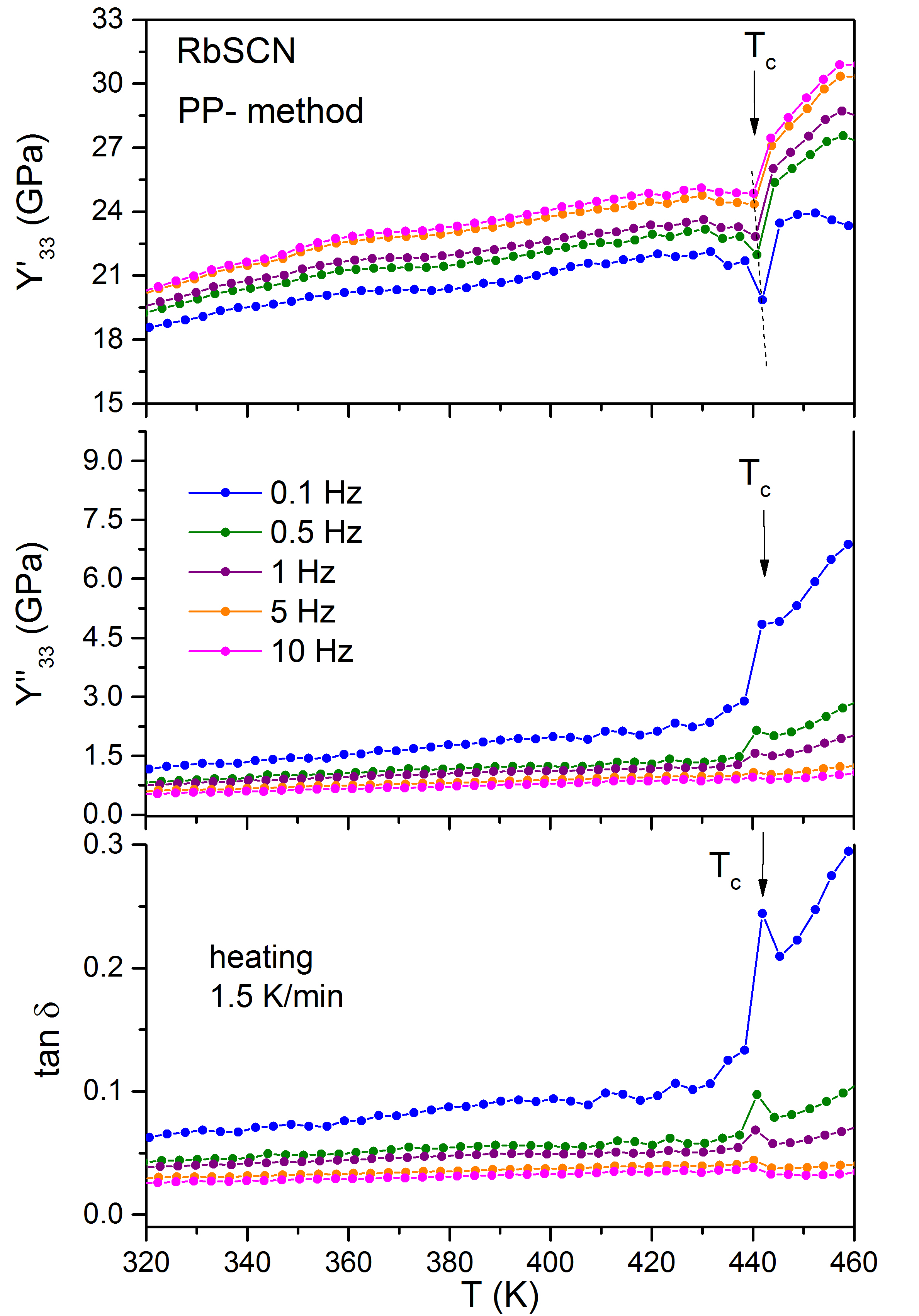}  
\caption{Temperature and frequency dependences of $Y'_{33}, Y''_{33}$ and $\tan \delta$ of RbSCN, measured in parallel plate mode with Diamond DMA (Perkin Elmer). Different frequency curves are shifted for clarity.}
\label{fig:Y33}
\end{figure}

The Young's modulus Y' and tan$\delta$ are determined by measuring the dynamic amplitude $u$ and phase shift $\delta$ between force and amplitude  in response to an applied dynamic force $F(t) = F_s +  F_d sin(\omega  t)$ using a Diamond DMA (Perkin Elmer). By subsequently applying a sinusoidal oscillation of various frequency, the dynamic elastic response is determined for a number of frequencies in a single temperature run.
Fig.\ref{fig:Y33} shows the elastic response with applied stress in c-direction at frequencies between 0.1 Hz and 10 Hz. At this direction no domain wall response is detected, as is expected for this geometry. The negative dip anomaly in $Y'_{33}$ as well as the peak in $Y''_{33}$ close to $T_c$ result from coupling terms of the type $\alpha \eta^2 \epsilon$ in the Landau free energy expansion. 

In a previous work \cite{Schranz1994} we have shown that at T$_c$ the decrease of the anomalies in  $Y'_{33}$ and  $Y''_{33}$ with increasing frequency (0.1 - 50 Hz) results from heat-diffusion, i.e. a crossover from isothermal to adiabatic elastic behaviour \cite{Schranz1994} when $\omega \tau_{th} < 1 \rightarrow \omega \tau_{th} > 1$. $\tau_{th}$ is the thermal relaxation time, which is of the order of 0.01~s.

\begin{figure}[h]
\centering
\includegraphics[scale=0.3]{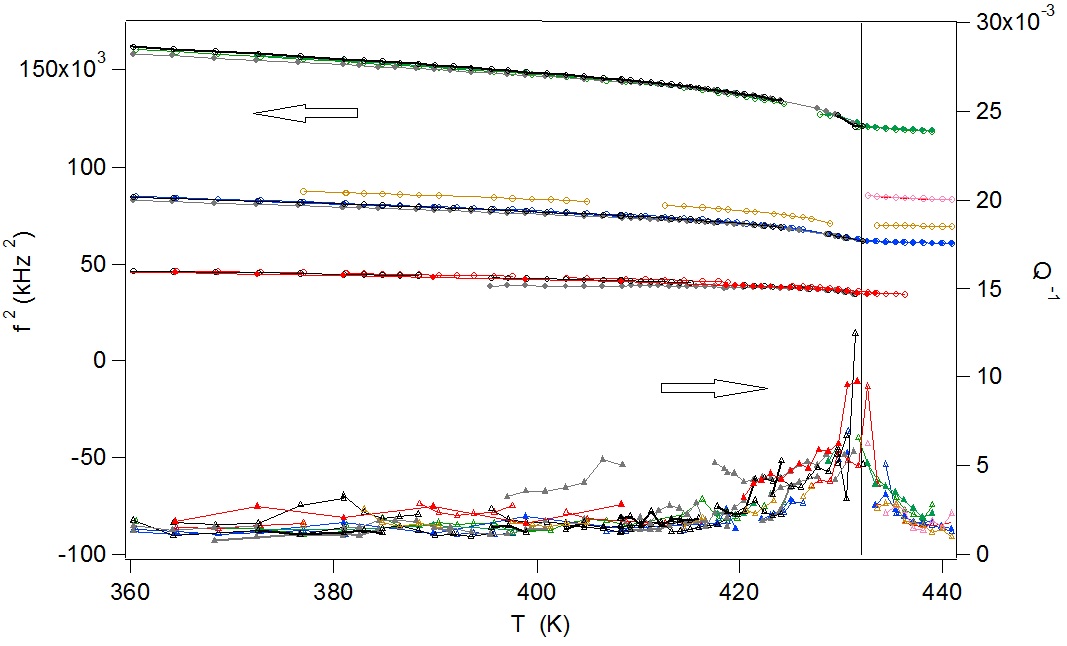}  
\caption{Temperature evolution of $f^2 \propto C_{el}$ and $Q^{-1} \propto tan \delta$ for some selected resonances in RUS spectra of RbSCN. All resonances show similar temperature dependencies.}
\label{fig:RUS}
\end{figure}

Fig.\ref{fig:RUS} displays results from resonant ultrasound spectroscopy (RUS) measurements, primary spectra were collected in the instrument described in Ref.\onlinecite{McKnight2008}. Individual peaks in the frequency range 100 - 600~kHz were fit with an asymmetric Lorentzian function to determine their frequency, $f$, and width at half maximum height, $\Delta f$. $f^2$ scales with different combinations of single crystal elastic constants, $C_{el}$. Acoustic loss scales with the inverse mechanical quality factor, $Q^{-1} = \Delta f/f$. 
The peaks were too broad to be measured at exactly the transition point but, otherwise, the results are quite clear. The transition is close to being continuous and there is a peak in $Q^{-1}$ at the transition point. There is no sign of any domain wall mobility and the elastic constants $C_{el} \propto f^2$ stiffen rather than showing the softening that would be expected if the order parameter was able to relax on a time scale of  $\approx10^{-6}$ s. In agreement with NMR results \cite{Blinc1995} this implies that $\tau_{\eta} > 10^{-6} s$, . 

\begin{figure}[h]
\centering
\includegraphics[scale=0.5]{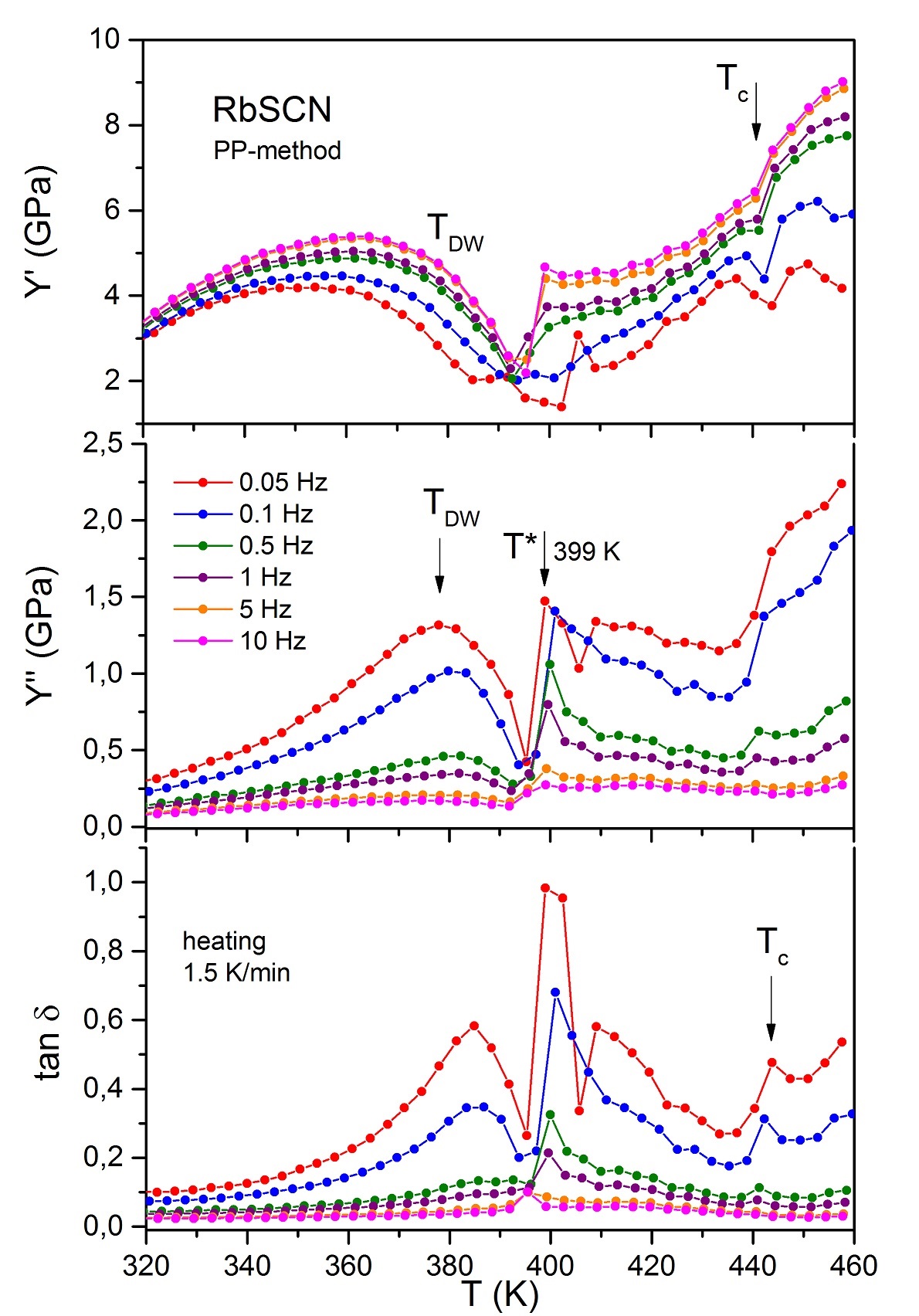}  
\caption{Temperature and frequency dependences of $Y'_{11}, Y''_{11}$ and $\tan \delta$ of RbSCN, measured during heating from room temperature in parallel plate mode with Diamond DMA (Perkin Elmer). The stress is applied in a-direction. {$T_{DW}$ marks the domain freezing temperature at $\omega \tau_{DW}(T) = 1$. $T^{\ast}$ is explained in the text.}}
\label{fig:Y11}
\end{figure}

\begin{figure}[h]
\centering
\includegraphics[scale=0.3]{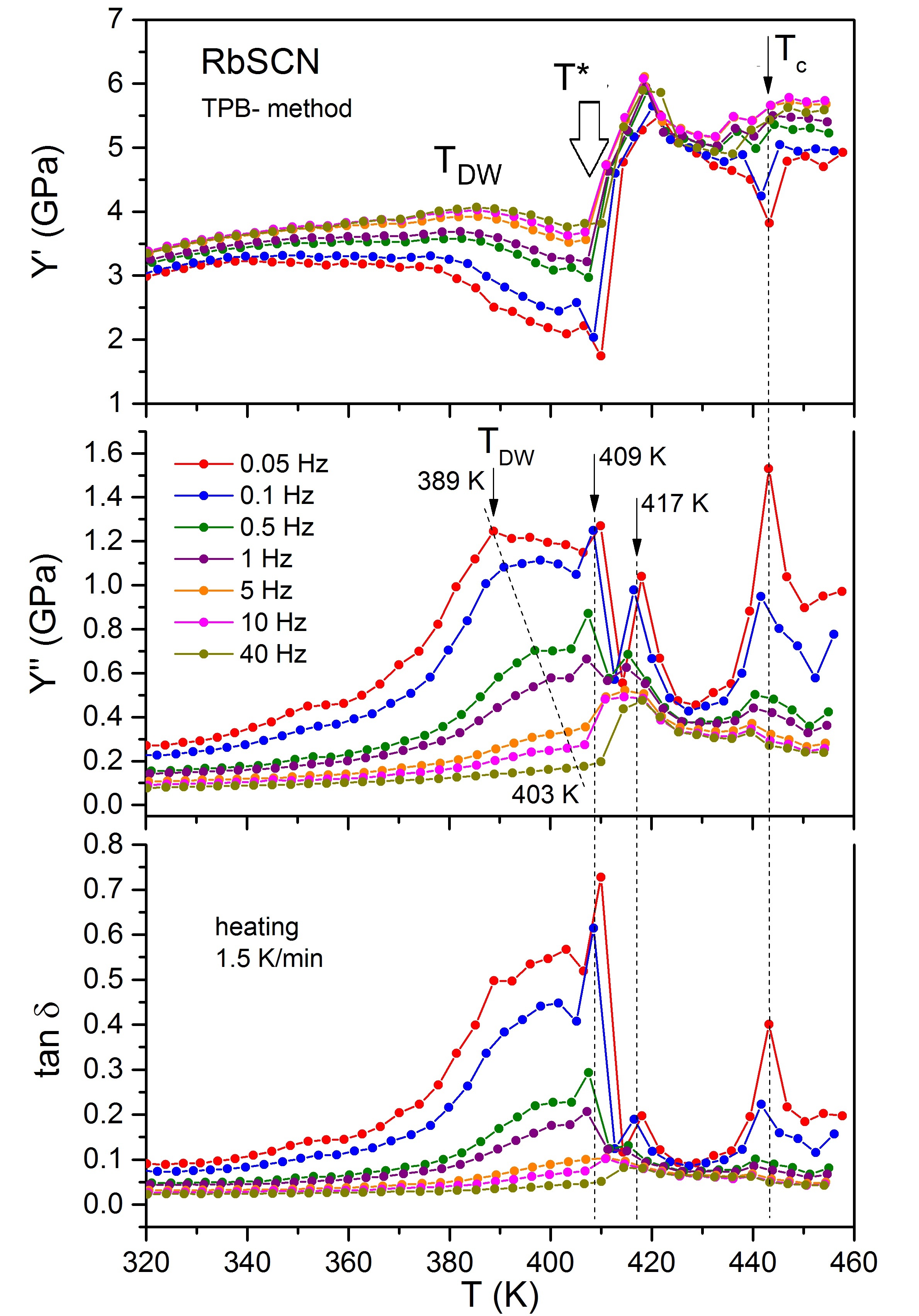}  
\caption{Temperature and frequency dependences of $Y'_{11}, Y''_{11}$ and $\tan \delta$ of RbSCN, measured in three point bending geometry with Diamond DMA. {With the applied force F=6.4~N and the sample dimensions, L=3 mm, b=2.78 mm, t=1.21 mm, the average (over sample thickness t) flexural stress $\sigma = 3 F L/4 b t^2 \rightarrow \sigma^{\ast}_2 \approx 4~MPa$, $T_2^{\ast} \approx 410~K$.}}
\label{fig:Y11 in TPB}
\end{figure}

In contrast to the Young's modulus in the c-direction, the a- and b-directions yield quite different behaviors at low frequencies (Figs.~\ref{fig:Y11} and \ref{fig:Y11 in TPB}).  

For stresses applied in a- or b-direction, the TBs contribute to the corresponding dynamic elastic behavior {partly}
quite similarly to what was observed in other multidomain ferroelastic materials  \cite{Schranz2011, Schranz2012}. 
{The most striking feature is the additional anomaly around a temperature  $T^{\ast}$ markedly lower than $T_c$. 
$Y'$ shows a discontinuous increase at $T^{\ast}$, which decreases with increasing frequency and is accompanied by a peak in $Y''$ whose magnitude also decreases with increasing frequency (Figs. \ref{fig:Y11} and \ref{fig:Y11 in TPB}). In contrast to the peak around T$_{DW}$, the one at T$^\ast$ does not shift with varying frequency. Such behavior is reminiscent of a first order phase transition. We will discuss its origin below.}\\
The peak at $T_{DW}$ shifts to lower temperature with decreasing frequency. It results from a decreasing mobility of TBs with decreasing temperature as the domain wall relaxation time $\tau_{DW}$ increases. The crossover from $\omega \tau_{DW} < 1$ to  $\omega \tau_{DW} > 1$  leads to a maximum in $Y''_{11}$ at $\omega \tau_{DW} = 1$ around 380~K.
Similar behavior has also been found recently in KSCN \cite{Soprunyuk2025}. However,
 {because of the absence of any $T^{\ast}$-anomaly}  in KSCN, the TB induced softening in $Y'$ between $T_{DW}$ and $T_c$ is clearly visible, whereas in RbSCN it is masked by processes occurring close to and above $T^{\ast} \approx 400~K$. 

\begin{figure}[h]
\centering
\includegraphics[scale=0.6]{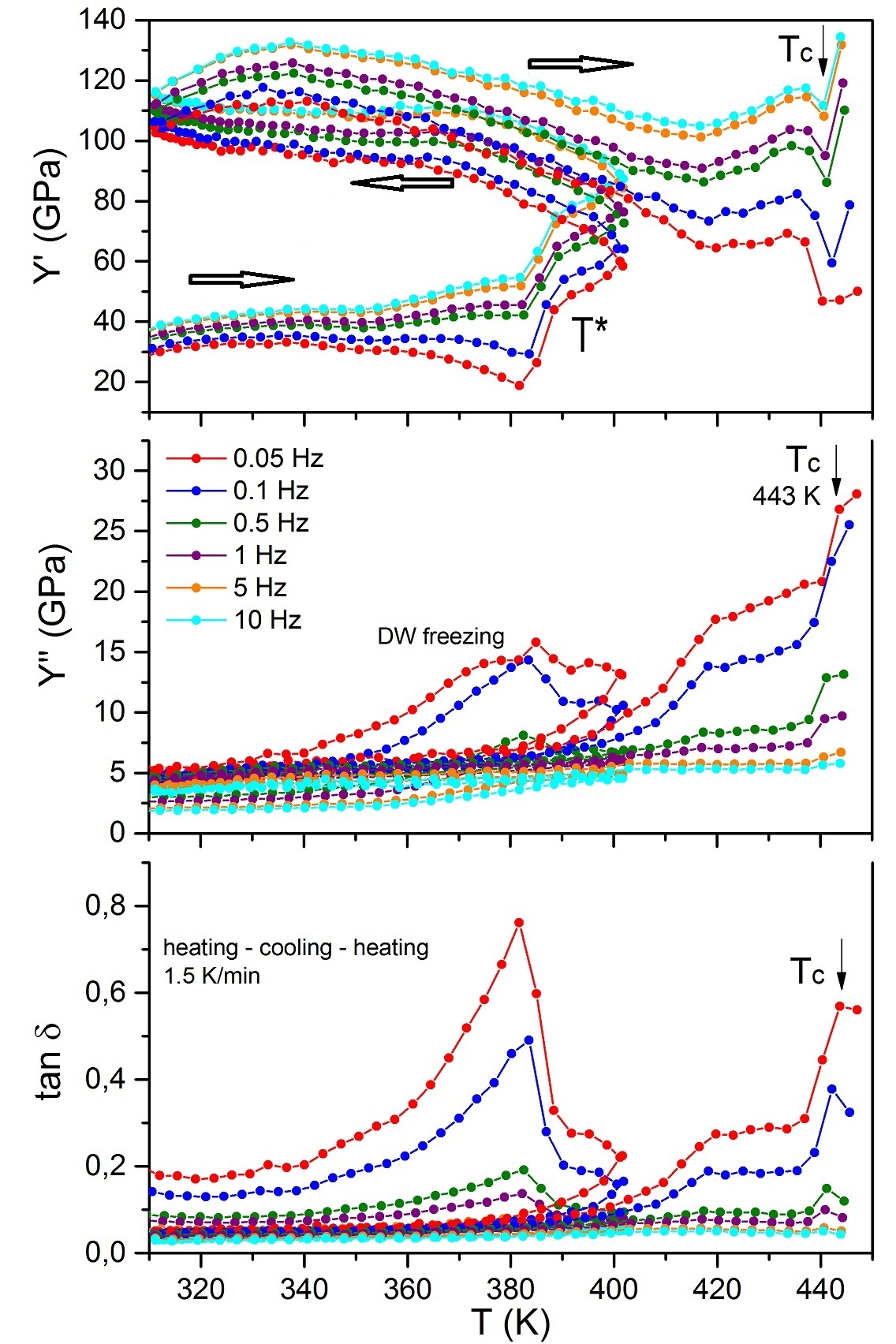}  
\caption{Temperature and frequency dependences of $Y'_{11}, Y''_{11}$ and $\tan \delta$ of RbSCN, measured in three point bending geometry with Diamond DMA for a sequence of heating above $T^{\ast} \approx 380~K$, followed by subsequent cooling to room temperature and heating again to a temperature above $T_c$.
{With the applied force F=9~N and the sample dimensions, L=2 mm, b=1.78 mm, t=0.51 mm, the average (over sample thickness t) flexural stress $\sigma = 3 F L/4 b t^2 \rightarrow \sigma^{\ast}_1\approx 30~MPa$, $T_1^{\ast} \approx 380~K$}.}
\label{fig:DWs removed}
\end{figure}

To get a clue about the anomalies around $T^{\ast}$ we heated the RbSCN crystal from room temperature to a temperature slightly above $T^{\ast}$, cooled it back to room temperature and heated again to temperatures above $T_c$. Fig.\ref{fig:DWs removed} shows a typical heating-cooling-heating run at a rate of 1.5~K/min. {Changing heating/cooling rates (1 - 5~K/min) did not change the behavior.}
One observes that subsequent cooling from $T>T^{\ast}$ removes all signatures of DW movement.
This is most clearly demonstrated by the absence of any peaks in $Y''$ or $\rm{tan} \delta$ after cooling from above $T^{\ast}$.
{Stopping the first heating run at $T<T^{\ast}$ and cooling back to room temperature, did not yield any anomaly, i.e. $Y'$ is identical with the first heating curve. }
This  suggests that the anomalies in RbSCN observed around $T^{\ast}$ are related to correlated (avalanche like) domain switching events in response to the applied stress in parallel plate and three point bending DMA measurements. 
{Interestingly enough, $T^{\ast} \approx 380~K$ in this case (Fig.\ref{fig:DWs removed}), while $T^{\ast} \approx 410~K$ in Fig.\ref{fig:Y11 in TPB}. Since both experiments differ in the applied forces, it implies that $T^{\ast}$ depends on the applied flexural stress.}

In the following section we will describe the various contributions to the dynamic elastic response of RbSCN in its monodomain and multidomain states.

\section{Discussion} \label{sec:Discussion}

\subsection{Monodomain response}  \label{sec:monodomain response}

Let us start to describe the different patterns shown in Fig.\ref{fig:Y33} - Fig.\ref{fig:DWs removed}.  
To calculate the intrinsic  (due to the PT) temperature dependencies of the elastic constants we use a compressible pseudospin model \cite{Schranz1989}.    
In the model, the anomaly in the longitudinal elastic constants is induced by the coupling terms between the order parameter $\eta$ and the strains $\epsilon_i$, (i=1,2,3), which are of the type $\alpha_i \eta^2 \epsilon_i$. The corresponding expression - sometimes called Landau-Khalatnikov (LK) contribution - can be written as \cite{Schranz1994b} $C_{ij} = C^\infty_{ij} + \Delta C^{LK}_{ij}$ with 

\begin{equation}
\label{eq:LK contribution}
\Delta C^{LK}_{ij} = - \frac{\alpha_i \alpha_j T_0}{V k_B T} \frac{\eta_s^2 (1-\eta_s^2)^2 (1+A \eta_s^2)}{T-T_0 (1-\eta_s^2) (1+A \eta_s^2)} 
\end{equation} 

where $T_0$ is the stability limit of the high temperature phase and the parameter $A$ measures the ratio between the elastic energy and the exchange energy $J = k_B T_0$. If $A < \frac{1}{3}$, the phase transition (PT) is of second order,  it is of first order for $A > \frac{1}{3}$ and $A =  \frac{1}{3}$ corresponds to a tricritical point at $T_0$. For RbSCN, a value of A $\approx$ 0.36 is reported \cite{Blinc1995}, thus being very close to a tricritical point. 
For KSCN, a value of A $\approx$ 0.5 was reported \cite{Schranz1994b,Schranz2005}, implying that the PT is of first order with $T_c - T_0 \approx 3 K$. \\
The equilibrium value $\eta_s$ of the order parameter (OP) is calculated from the relation

\begin{equation}
\label{eq:OP}
\eta_s(T) = \rm{tanh}\left[ \frac{T_0}{T}(\eta_s + A \eta_s^3)\right]
\end{equation} 

Eq.(\ref{eq:LK contribution}) holds for $\omega \tau_{\eta}<1$ and $\omega \tau_{th}<1$, where $\tau_{\eta}$ is the order parameter relaxation time, which determines the time scale of OP fluctuations and $\tau_{th}$ is the thermal relaxation time, characterizing the temperature exchange between different regions in the crystal.   
The observed dip anomalies in the real part and the peak at $T_c$ in the imaginary part of the Young's modulus shown in Figs.\ref{fig:Y33}, \ref{fig:Y11} and \ref{fig:Y11 in TPB} are typical for the LK-contribution (\ref{eq:LK contribution}).  The decrease of these anomalies with increasing frequency implies a crossover from $\omega \tau < 1 \rightarrow \omega \tau > 1$. The absence of any dip anomalies in $f^2 \propto C_{el}$ for RUS measurements (Fig.\ref{fig:RUS}) demonstrates that the order parameter cannot relax on the time scale of the strain variations (some $10^5$ Hz), implying $\tau_{\eta} > 1/\omega \approx 10^{-6}s$. In previous work we have shown \cite{Schranz2005, Soprunyuk2025} that in such a case the anomalies can be fitted by $C_{el} \propto \eta_s^2$, originating from additional $\eta^2 \varepsilon^2$ terms in the corresponding Landau free energy expansion.

Quite interestingly, $Q^{-1}$ data (Fig.\ref{fig:RUS}) display peaks at all frequencies even above $T_c$, whereas no such precursor effects are found in the resonance frequencies which are proportional to the elastic constants. 
Such unusual behavior was also found recently in KSCN \cite{Soprunyuk2025}, and explained in terms of order parameter fluctuations of rather large time scale ($\tau_{\eta} > 10^{-6}$s).

\subsection{Effects of twin boundary motion}  \label{sec:DW motion effects}

From Figures \ref{fig:Y11} and \ref{fig:Y11 in TPB} it is obvious that in contrast to the c-direction (Fig.\ref{fig:Y33}),  Young's moduli in a- and b-directions are dominated by TB motion. TB motion leads usually to superelastic softening in $Y'$, accompanied by a broad peak in $Y''$, whose maximum at the temperature $T_{DW}$ (at $\omega \tau_{DW} = 1$) shifts to higher temperatures with increasing frequency of the applied stress \cite{Schranz2011, Schranz2012, Harrison2004,Puchberger2016, Kityk2000}. Such behavior is indeed found for RbSCN (Figs. \ref{fig:Y11} and \ref{fig:Y11 in TPB}). However, it is interrupted by a discontinuous increase of $Y'$ at $T^{\ast}$,  which is accompanied by a peak in $Y''$. 
{The temperature $T^{\ast}$ of the anomalies  does not shift with frequency, but shifts to lower temperature with increasing stress. In the following we will substantiate that the anomalies at $T^{\ast}$ are related to domain switching.}

To calculate the TB contributions to the dynamic elastic response we follow the theoretical elaborations of Refs.\onlinecite{Tagantsev2010, Sidorkin1997, Sidorkin2012, Darinskii1987}.
The application of an external macroscopic stress $\sigma$ results in a pressure $ \sigma \varepsilon_s$ applied to TBs existing in the ferroelastic state. {This results from the strain contrast $\varepsilon_s = \varepsilon_1^h - \varepsilon_2^h$ between two adjacent ferroelastic domains $1_1$ and $2_1$ and will finally result in a motion of the ferroelastic DWs. For adjacent antiphase boundaries (APBs) the homogeneous strains are equal on both sides, implying $\varepsilon_s = 0$, which implies that APBs cannot be moved by  external stress.}

Quite generally, the movement of DWs in external fields can be hindered due to pinning by various obstacles, i.e. point defects, dislocations, surface, neighbouring TBs or antiphase boundaries, etc. Here we will consider the case of \textit{lattice pinning} (Peierls pinning) as TBs in RbSCN and KSCN turn out to be very thin. 
Depending on the relation between the external stress and the strength of pinning, one can roughly distinguish two regimes of DW motion, {i.e. thermally activated and non-activated regime.}  
Let us recall an explanation of the two regimes \cite{Tagantsev2010, Sidorkin1997}.

\begin{figure}[h]
\includegraphics[scale=0.35]{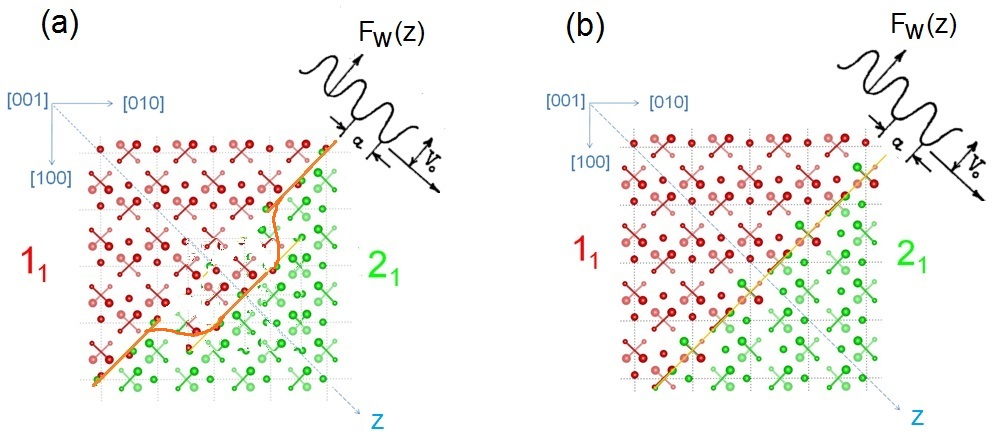}  
\caption{Sketch of the movement of a TB segment trapped in the Peierls potential, forming a nucleus of the inverse domain.
(a) a TB segment jumped over the barrier, (b) the TB is at the saddle point.}
\label{fig:Peierls potential}
\end{figure}

For very thin DWs the surface energy density $F_w(z)$ of a domain wall depends on the position $z$ of the wall center with respect to the sites of the crystal lattice (Figure \ref{fig:Peierls potential}).

It is given as \cite{Sidorkin1997}

\begin{equation}
\label{eq:lattice pinning}
F_w(z) = F_w(0) + \frac{V_0}{2} \rm{cos} \left( \frac{2 \pi z}{a} \right)
\end{equation}

{where $V_0$ is related to the pinning barrier and reads}

\begin{equation}
\label{eq:Peierls potential}
V_0 = 8 \pi^4 F_w(0) \left( \frac{w}{2a} \right)^3 \rm{exp} \left(-\pi^2 \frac{w}{2a} \right)
\end{equation}
  
$w$ is the DW thickness and $F_w(0)$ is the TB energy density as calculated e.g. from a continuous Landau-Ginzburg model \cite{Rychetsky1994} at $z=0$.

{The exponential dependence of $V_0$ on the TB thickness $w$ leads to a very pronounced temperature dependence of a number of physical quantities.} 
For narrow TBs with $w \approx a$, the presence of the pinning barriers prevents their motion as a whole at relatively weak external stresses, as this would require to overcome energy barriers of the order $E_B = S_0 V_0 \gg k_B T$ ($S_0$ is the area of the TB). 
In this so called \textit{thermally activated regime} $\sigma < \sigma_c$, only small segments of the TB can perform thermally activated jumps forming nuclei of the reverse domain (Fig.\ref{fig:Peierls potential}). The crossover from the \textit{activated} to the \textit{non-activated} regime \cite{Sidorkin1997} occurs at a critical stress of about 

\begin{equation}
\label{eq:sigmac}
\sigma_c \approx \frac{V_0}{\varepsilon_s a}
\end{equation}

where $\varepsilon_s = \varepsilon_1^h - \varepsilon_2^h$ { is the strain contrast between adjacent ferroelastic domains.  As we show in the SI (Table I), the homogeneous spontaneous strains in adjacent domains are, $\varepsilon_1^h=k_1 \eta_s^2$ and $\varepsilon_2^h=k_2 \eta_s^2$, respectively. As a result, $\varepsilon_s = (k_1-k_2)\eta_s^2 = 0.003~\eta_s^2(T)$.}

The critical stress is strongly temperature dependent (Fig.\ref{fig:sigmacrit}) through $V_0$ (Eq. \ref{eq:Peierls potential}), which depends exponentially on the TB thickness $w(T)$. In Ref.\onlinecite{Rychetsky1994} we have calculated the TB thickness and corresponding energies for KSCN crystals, using a Landau-Ginzburg free energy expansion in the restricted temperature range $T_c - T \approx 5~K$. To extend the temperature range and to calculate these properties also for RbSCN, we use here a modified version of the compressible pseudospin model \cite{Schranz1989} by appending the free energy with gradient terms. The calculations are presented in the SI.

\begin{figure}[h]
\includegraphics[scale=0.32]{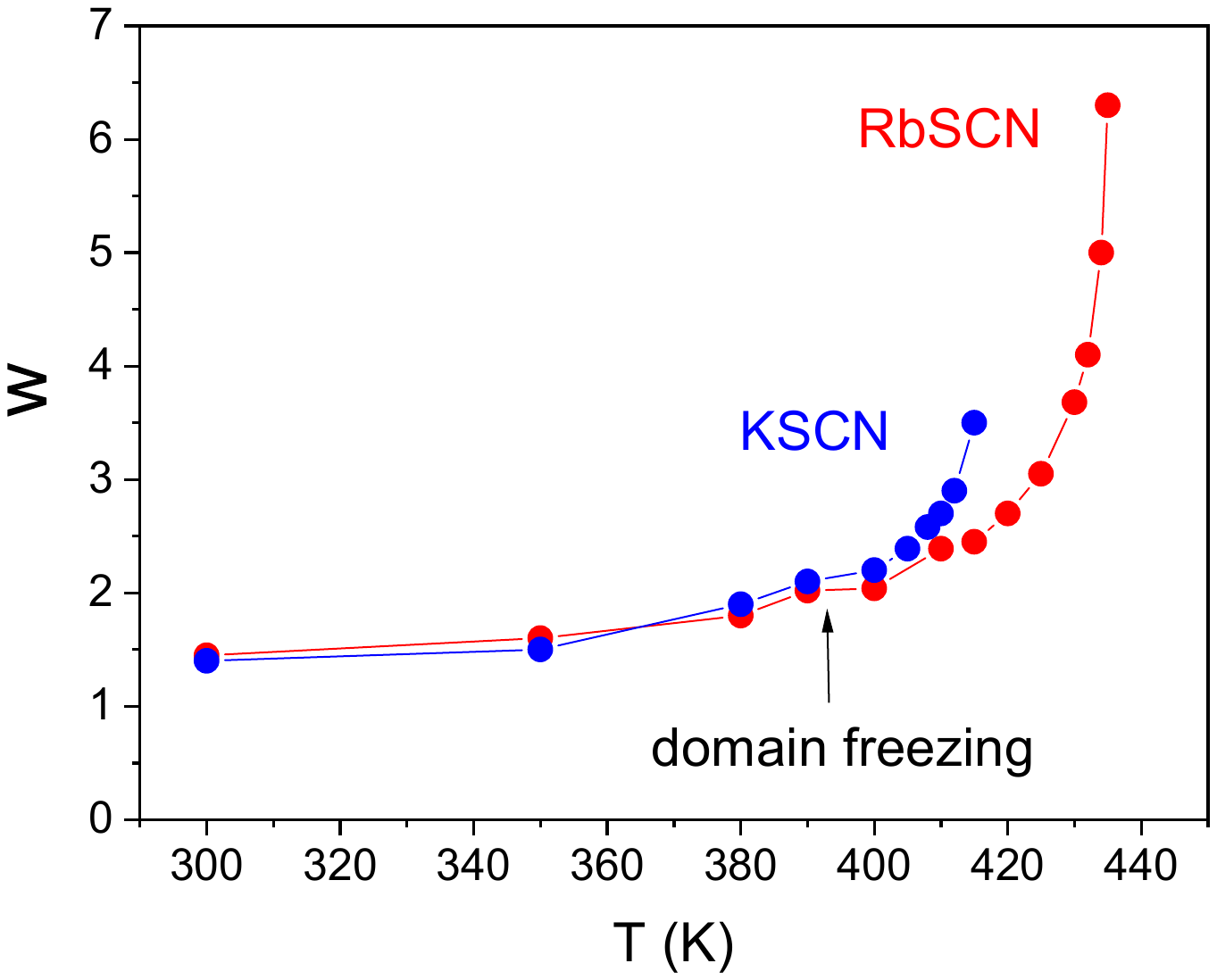}  
\caption{Temperature dependence of the width {(w in multiples of lattice constants)} of twin boundaries of RbSCN and KSCN. Details of the calculation are given in the SI.}
\label{fig:TB width}
\end{figure}

\begin{figure}[h]
\begin{center}
\includegraphics[scale=0.3]{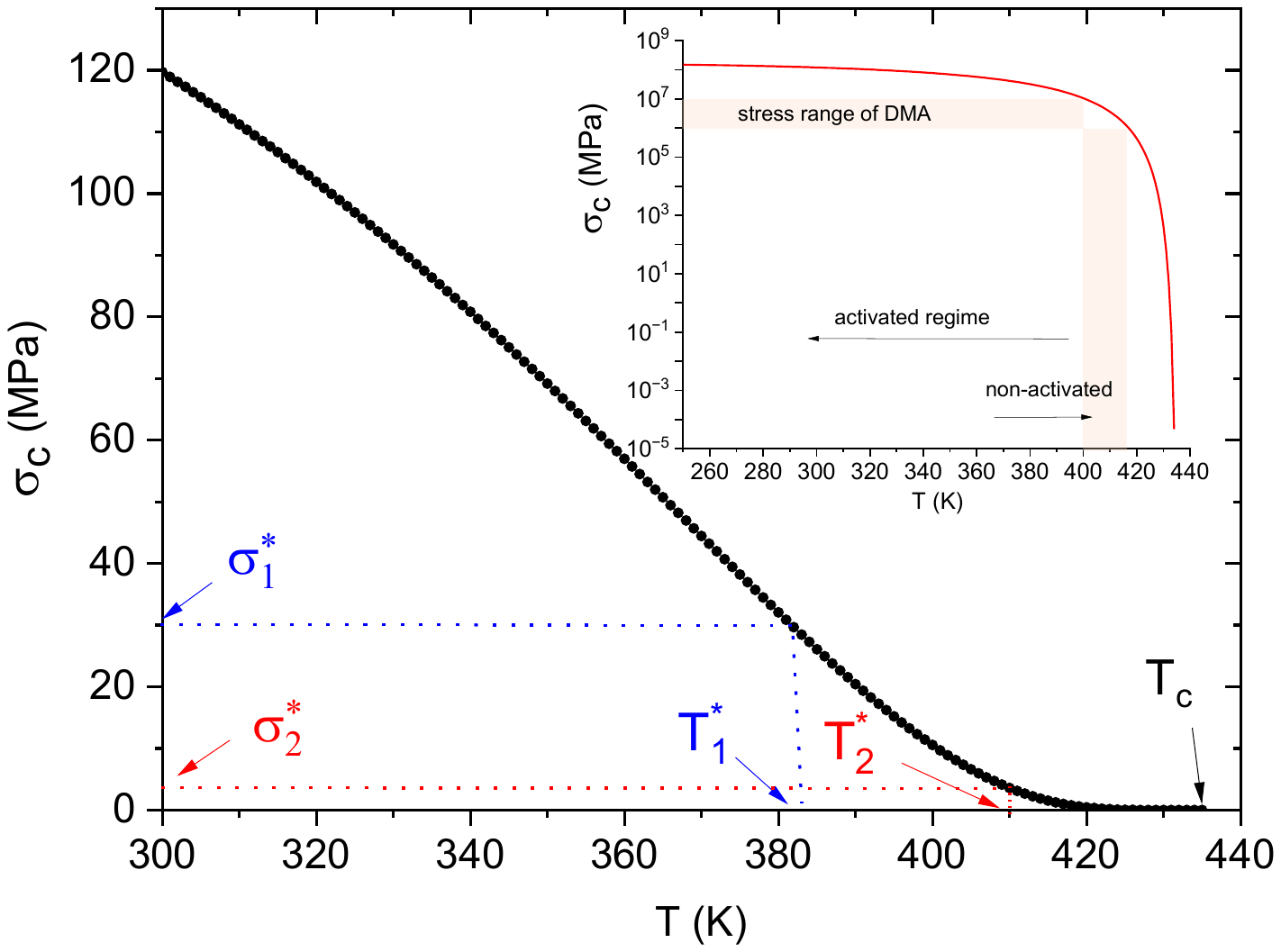}  
\caption{Temperature dependence of the critical stress, which is necessary to unpin TBs in RbSCN.
{$\sigma^{\ast}_1, T^{\ast}_1$ corresponds to Fig.\ref{fig:DWs removed}
 and $\sigma^{\ast}_2, T^{\ast}_2$ corresponds to Fig.\ref{fig:Y11 in TPB}.}}
\label{fig:sigmacrit}
\end{center}
\end{figure}

Fig.\ref{fig:TB width}  shows the resulting temperature dependence of the twin boundary (TB) widths of RbSCN and KSCN as calculated in the SI. Interestingly, the region of DW freezing appears for both crystals at a temperature where the TB width approaches a constant value of about 1 - 2 lattice constants. A similar behavior has been found previously for domain freezing in KDP (cf. Fig. 4 of Ref.\onlinecite{Sidorkin1997}).

Inserting $w(T)$ and $\varepsilon_s(T)$  from the SI into Eq.(\ref{eq:sigmac}) we arrive at the temperature dependence of the critical stress $\sigma_c(T)$, shown in Fig.\ref{fig:sigmacrit}. Between room temperature and about 400~K, the critical stress $\sigma_c$ varies only slightly, while between ca. 400 - 435~K it decreases by 10 orders of magnitude, due to the exponential dependence on $w(T)$.\\

In the present DMA experiments the applied stress $\sigma$ was in the range of { $3 - 30$ MPa}. Thus, $\sigma < \sigma_c$ for temperatures below approximately 380~K. {The following discussion relates to this activated temperature regime, in which only segments of TBs can be moved (thermally activated), thereby contributing to a TB induced decrease of $Y'$ and corresponding damping peaks in  $Y''$ (and $\rm{tan} \delta$) below $T^{\ast}$ (Fig. \ref{fig:Y11} and Fig. \ref{fig:Y11 in TPB})}. According to  \cite{Sidorkin1997} this can be written as

\begin{equation}
\label{eq:DW contribution to elastic modulus}
Y(\omega) = Y^{\infty} - \frac{\Delta Y^{DW}}{1+i \omega \tau_{DW}}
\end{equation}

{$\tau_{DW}$ is the characteristic time that it takes for a DW to settle into a new equilibrium position after being disturbed by an external field (stress in the present case).}
The maxima of the damping peaks occur at a temperature $T_{DW}$, where $\omega \tau_{DW} = 1$. 
The shift of the maxima of the damping peaks with varying frequency (Figs. \ref{fig:Y11} and \ref{fig:Y11 in TPB}) is used to determine the {DW relaxation time  $\tau_{DW}(T)$.
Adopting a simple Debye model we can write}

{
\begin{equation}
\label{eq:tau activated}
\tau_{DW}(T) = \tau_0 e^{\Pi^{\ast}/k_BT}
\end{equation}}

{where $\tau_0$ is an effective attempt frequency and $\Pi^{\ast}$ is the energy barrier for forming a critical nucleus  of the switched domain segment (Fig.\ref{fig:Peierls potential} a). In the simple approximation of a spherical nucleus of radius R, one obtains for the energy of the nucleus of the other domain \cite{Sidorkin2012}}

{
\begin{equation}
\label{eq:energy of nucleus}
\Pi = \frac{2 a}{\pi} \sqrt{2 F_w(0)V_0} 2 \pi R - 2 \varepsilon_s \sigma R^2 \pi a
\end{equation}}

{The positive terms depending linearly on R originates from lattice pinning \cite{Darinskii1987}. The negative term quadratic in R is the energy gain due to applied stress.
Minimizing $\Pi(R)$ with with respect to $R$ (see also SI) one obtains the energy barrier $\Pi^{\ast}$ for forming a critical nucleus of radius $R^{\ast}$ as}

{
\begin{equation}
\label{eq:critical energy of nucleus}
\Pi^{\ast} = \frac{4 a F_w(0)V_0}{\pi  \varepsilon_s \sigma}
\end{equation}}

{Inserting Eq.(\ref{eq:critical energy of nucleus}) into Eq.(\ref{eq:tau activated}) we get}

{
\begin{equation}
\label{eq:DW relaxation time}
\tau_{DW} = \tau_0 e^{ \frac{4 a F_w(0)V_0}{\pi  \varepsilon_s \sigma k_B T}}
\end{equation}}

{Although we started from a simple Arrhenius relaxation time Eq.(\ref{eq:tau activated})}, according to Eq.(\ref{eq:DW relaxation time}) the DW relaxation time follows a double exponential temperature dependence, since $V_0$ is already exponentially dependent on T. 

\begin{figure}[h]
\includegraphics[scale=0.3]{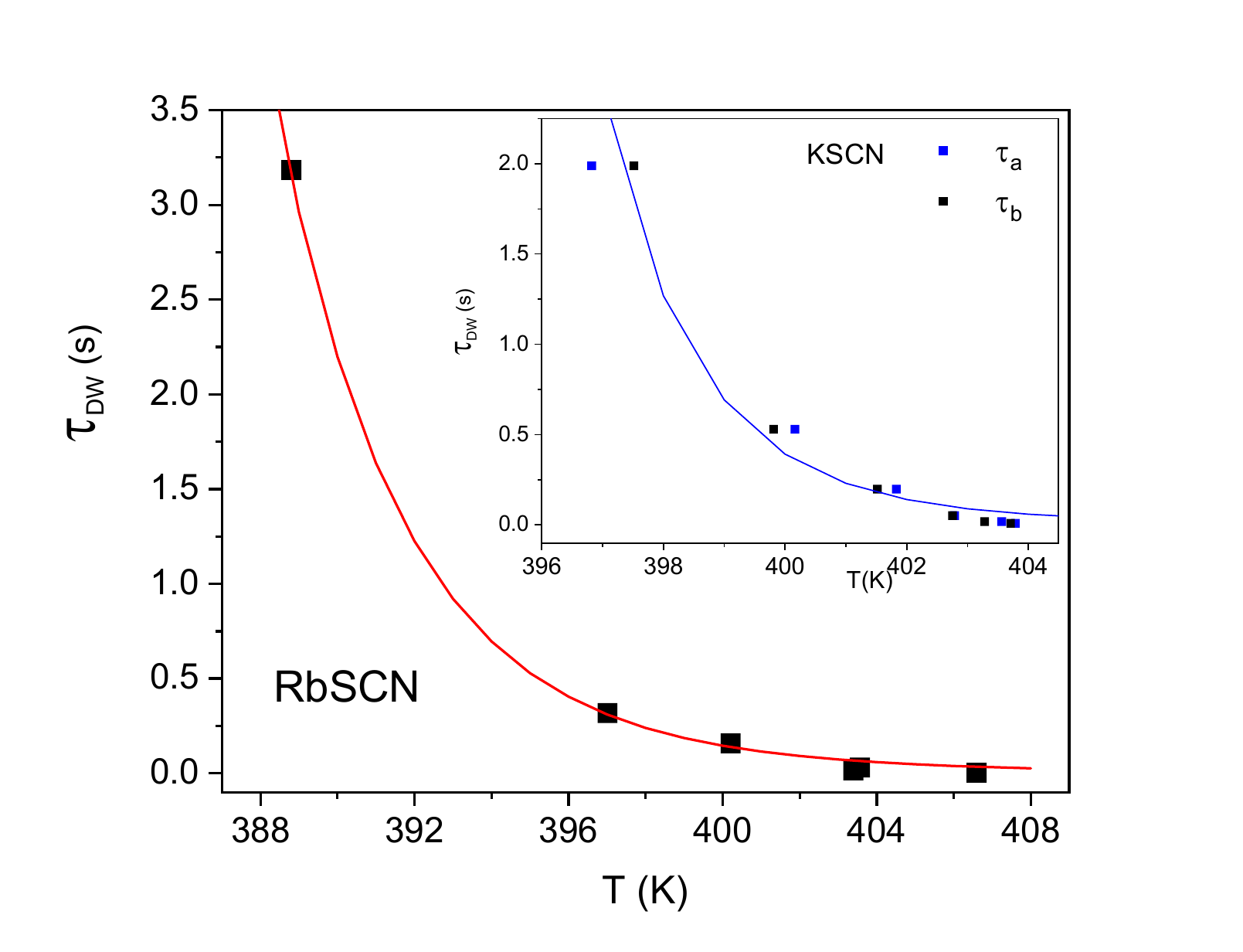}  
\caption{Temperature dependence of the DW relaxation times of RbSCN and (inset) KSCN. Points are determined from the peak shifts in $Y''$ and the lines are fits using Eq.~(\ref{eq:DW relaxation time}) with $\delta$ calculated in the SI, and the applied stress is $\sigma = 2 MPa$.}
\label{fig:DW relaxation time}
\end{figure}

Fig.\ref{fig:DW relaxation time} shows the temperature dependences of the DW relaxation times of RbSCN and KSCN determined from the peaks in $Y''$. The double exponential law of Eq.(\ref{eq:DW relaxation time}) fits the data perfectly. This agreement with the data is even more remarkable, as Eq.(\ref{eq:DW relaxation time}) leaves, instead of the effective prefactor ($\tau_0 \approx 3 \times 10^{-2}~s$), no free parameters. All quantities entering Eq.(\ref{eq:DW relaxation time}), i.e. $F_w(0), V_0, \varepsilon_s$ are calculated from the compressible pseudospin model (cf. SI).

The mean number of other domain nuclei $N$ appearing per time unit at the unit area of a TB increases exponentially with decreasing activation energy $\Pi^{\ast}$ and thus takes the form \cite{Sidorkin2012} (c.f. SI) 

\begin{equation}
\label{eq:number of nuclei}
N(T) \propto e^{-\Pi^{\ast}/ k_B T}
\end{equation}

\begin{figure}[h]
\includegraphics[scale=0.3]{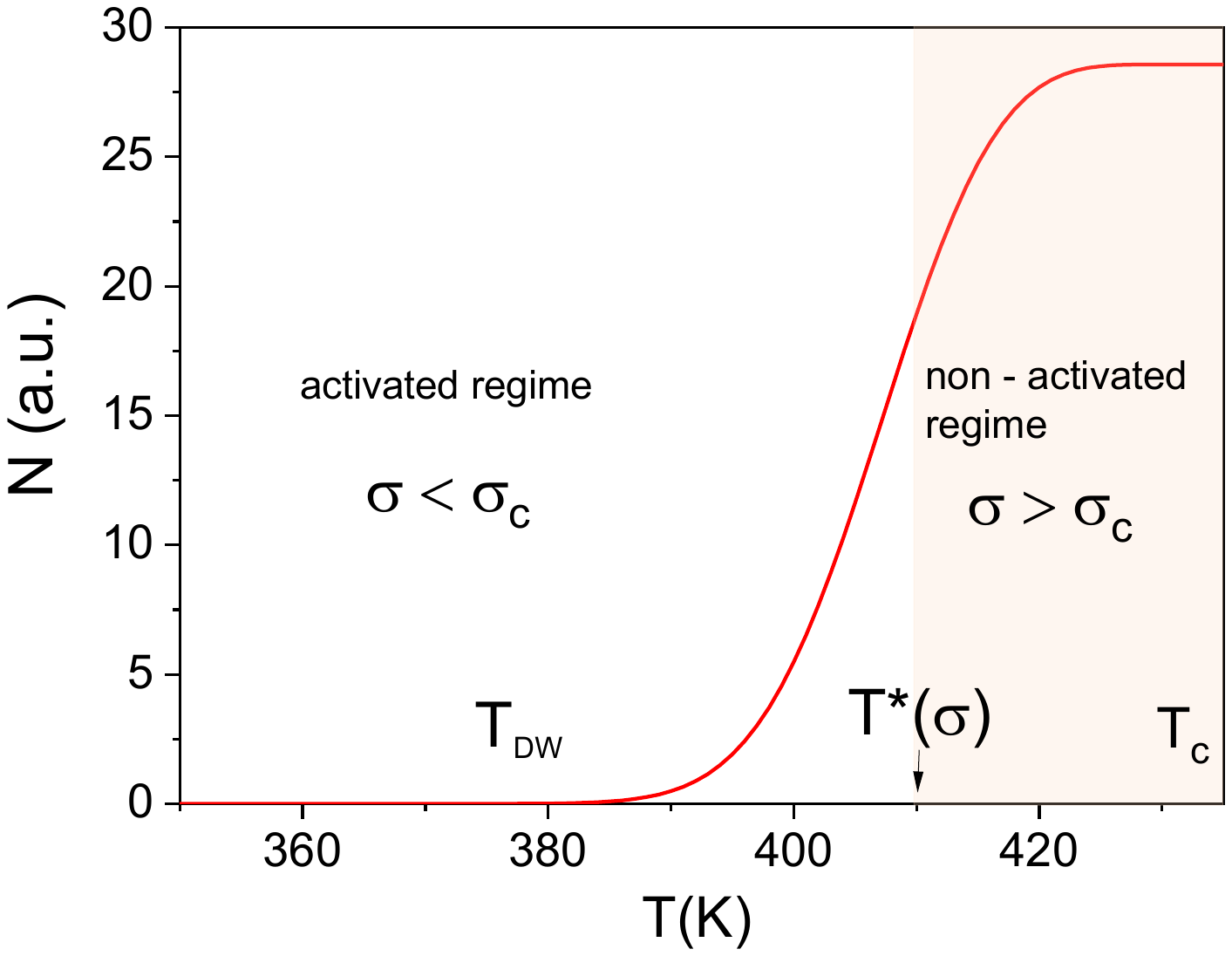}  
\caption{Temperature dependence of the mean number $N$ of inverse domain nuclei appearing per time unit of the unit area of a twin boundary in RbSCN, calculated from Eq.(\ref{eq:number of nuclei}) with $\sigma = 4 MPa$.}
\label{fig:number of nuclei}
\end{figure}

{With increasing temperature, the number $N(T)$ of nuclei increases, due to the decreasing Peierls potential (Eqs.(\ref{eq:Peierls potential}) and (\ref{eq:critical energy of nucleus})). In the temperature range above ca. 380~K, $N(T)$ starts to increase substantially, simultaneously with the increase of the DW width (Fig.\ref{fig:TB width}).   
At the temperature $T^{\ast}$ one approaches the region where the applied stress $\sigma$ starts to overcome the critical pinning stess $\sigma_c$. In the range $\sigma > \sigma_c$ (the \textit{non-activated regime})  an increasing number of   
TB segments in RbSCN become increasingly mobile (the DW relaxation time $\tau_{DW}$ is very small, Fig.(\ref{fig:DW relaxation time})) and these TB movements can trigger further events (Figure \ref{fig:number of nuclei}) until large parts of the crystal have collectively switched.\\
As seen in Figures  \ref{fig:Y11} and \ref{fig:Y11 in TPB}, this switching results in a discontinuous increase in $Y'$ and a corresponding peak in  $Y''$ and $\rm{tan} \delta$ at $T^{\ast}$. 
In contrast to the damping peaks below $T^{\ast}$, these peaks do not shift with frequency. However, they do shift with applied stress. In the parallel plate experiment of Fig.\ref{fig:Y11} the applied force was approximately 8~N, whereas $F \approx 6~N$ in the three point bending experiment shown in Fig.\ref{fig:Y11 in TPB}, resulting in $T^{\ast}(\sigma) \approx 400~K$ (Fig.\ref{fig:Y11}) vs. (410~K) (Fig.\ref{fig:Y11 in TPB}). In Fig.\ref{fig:sigmacrit} we show that the experimentally observed decrease of $T^{\ast}$ with increasing stress fits perfectly with the calculation of the critical pinning stress $\sigma_c(T)$ using Eq.(\ref{eq:sigmac}). 
In the experiment shown in Fig.\ref{fig:DWs removed} the initial heating run was performed to a temperature slightly above $T^{\ast} \approx 380~K$, followed by cooling back to room temperature and then heating again by crossing $T_c$. After cooling from $T \geqslant T^{\ast}$ all DW relaxation peaks disappeared, implying that above $T^{\ast} \approx 380~K$, the crystal is indeed in a switched state.}

\section{Summary and conclusion}   \label{sec:Conclusion}

The present dynamic elastic measurements of RbSCN in a broad range of temperatures (300 - 460 K) and frequencies (0.05 Hz - 40 Hz, and 100 - 600 kHz) {and at various applied stresses} 
 yield insights into the rich dynamical behavior of twin boundaries. Ferroelastic domains appear either on cooling below $T_c \approx 435~K$, and/or are already present in as grown samples of RbSCN. The corresponding twin boundaries are oriented along $(110)$ or $(\bar{1}10)$ planes.   
Due to the strain contrast between adjacent ferroelastic domains, the corresponding domain walls are mobile under dynamic stress applied in the a - or b - direction, which leads to superelastic softening in the corresponding Young's moduli $Y'_{11}$ and $Y'_{22}$ below $T_c$. The softening is accompanied by frequency dependent peaks in $Y''_{11}$ and $Y''_{22}$, shifting to lower temperatures with decreasing frequencies. The strong shift of these peaks - with maxima at $\omega \tau_{DW} =1$ - is reflected in a sharp increase of the domain wall relaxation time, $\tau_{DW}$,  with decreasing temperature. A first attempt to fit the relaxation time using an Arrhenius law leads to meaningless values of the pre-exponential factor and the activation energy. A Vogel-Fulcher law, with $\tau_{DW}=\tau_0~exp[E/k_B(T-T_{VF})]$, fitted the data very well with $\tau_0 \approx 10^{-5}~s$, $E \approx 0.037~eV$ and $T_{VF} \approx 354~K$. In a recent work on KSCN \cite{Soprunyuk2025} we fitted the DW relaxation time also by a Vogel-Fulcher law and obtained similar values, i.e. $\tau_0 \approx 10^{-7}~s$, $E \approx 0.035~eV$ and $T_{VF} \approx 368~K$. 
These values are quite sound and may be related to jamming \cite{Salje2014, Salje2015} as one of the prime mechanisms that generates glass like states with glass freezing in crystals containing such high twin wall densities. 
The reason for Vogel-Fulcher-distributed dynamics may be due to the heterogeneous structure of the domain glass: nucleation and propagation of domain walls see different energy barriers.

However, since the Vogel-Fulcher law is purely phenomenological, to get a deeper insight into the microscopic mechanisms of TB motion in RbSCN, we adopted a Landau-Ginzburg type theory of \textit{thin} domain walls \cite{Sidorkin1997, Sidorkin2012}. Our calculations, based on a compressible pseudospin model, show that TBs in RbSCN and also in KSCN are indeed very thin (Fig.\ref{fig:TB width}), with a nearly constant thickness of about 1-2 lattice constants from room temperature up to about 400~K. In this temperature range, the movement of the TBs in response to the applied dynamic stress occurs via thermally activated jumps of TB segments over energy barriers that are determined by the Peierls energy. This activation energy decreases exponentially with increasing TB width, which leads to a double exponential decrease of the DW relaxation time $\tau_{DW}$ with increasing temperature. Quite remarkably, Eq.(\ref{eq:DW relaxation time}) fits the experimental data on $\tau_{DW}$ perfectly (Fig.\ref{fig:DW relaxation time}) without any free parameter. The movement of the thin TBs in the lattice relief occurs by thermally activated jumps up to a critical stress $\sigma_c$, which is also exponentially decreasing with increasing temperature (Fig.\ref{fig:sigmacrit}).

Around 400~K the critical stress $\sigma_c$ starts to decrease by orders of magnitude due to an increase of the TB width. 
In this temperature region, the number of critical nuclei of the inverse domain increase enormously and the applied stress starts to exceed $\sigma_c$. In this non-activated regime, where the TBs are no longer pinned by the lattice potential,         
there appears to be a discontinuous increase of the Young's moduli in a- and b-directions around {$T^{\ast}(\sigma) \approx 380 - 410~K$}, accompanied by a peak in $Y''$ and tan$\delta$. Heating RbSCN above about 380 - 410~K and cooling down again, removes all signs of DW motion. This shows that, around the temperature $T^{\ast}(\sigma)$ , the TB segments become indeed unpinned ($\sigma > \sigma_c$) and their mobility gets very high ($\tau_{DW} < 0.1~s$), leading to a correlated avalanche-like switching of ferroelastic domains.\\
{The question arises, why similar domain switching at $T^{\ast}$ was not observed in recent experiments on the related crystals KSCN \cite{Soprunyuk2025}, as there are no fundamental differences between RbSCN and KSCN. In both crystals the domain dynamics is mainly governed by thermally activated jumps of pinned domain wall segments over energy barriers that depend exponentially on DW thickness. Similar to RbSCN, the DW relaxation time of KSCN can also be fitted with a double exponential temperature dependence (inset of Fig.\ref{fig:DW relaxation time}). However, unlike for RbSCN, we did not observe any stress induced domain switching for KSCN  for the following reasons: 
The stresses in three point bending measurements of KSCN were generally lower (2 - 4 MPa) compared to RbSCN (4 - 30 MPa), and  the critical stress needed to unpin the TBs in KSCN is a bit larger and approaches zero less steeply due to the fact that TBs in KSCN are thinner as those in RbSCN (Fig.\ref{fig:TB width}). This shifts $T^{\ast}(\sigma)$ for KSCN very close to $T_c$, i.e. $T_c - T^{\ast} \approx 5~K$. At this temperature close to $T_c$, the effect of superelastic softening in Y' is already too small to allow for the observation of any detectable difference between a multidomain and a single domain response.}

Summarizing, we think that despite of the simplifications made here, the present work gives valuable insights into the microscopic mechanisms of TB motion. {In particular, the twin boundary thickness, its structure and temperature dependence seem to be key to explaining domain dynamics and domain freezing. Further work is needed to see to what extent the present physical insights can also be applied to other ferroelastic materials.}\\

\noindent \textbf{Acknowledgments} 
This research was funded by the Austrian Science Fund (FWF) 10.55776/PIN2246224 and 10.55776/PAT2916124. 
RUS facilities in Cambridge were funded by the Natural Environment Research Council of Great Britain (Nos. NE/B505738/1 and NE/F017081/1). Part of this work was supported by the Czech Science Foundation (GACR), project No. 25-18870L and by the e-INFRA CZ project (ID: 90254).\\

\end{document}